\documentclass[12pt]{article}
\usepackage[latin1]{inputenc}
\usepackage[dvips]{graphicx,epsfig,color}
\usepackage{wrapfig,rotating}
\usepackage{amssymb,amsmath,array}

\def\beq{\begin{equation}}
\def\eeq#1{\label{#1}\end{equation}}
\def\eeqn{\end{equation}}

\def\beqa{\begin{eqnarray}}
\def\eeqa#1{\label{#1}\end{eqnarray}}
\def\eeqan{\end{eqnarray}}

\let\bar=\overbar

\def\Dslash{\not{\hbox{\kern-4pt $D$}}}
\def\dslash{\not{\hbox{\kern-2pt $\del$}}}

\def\msb{{\bar{\ssstyle M \kern -1pt S}}}

\def\Title#1{\begin{center} {\Large {\bf #1} } \end{center}}

\begin{document}

\Title{Longitudinal Structure Function Measurements from 
  HERA\footnote{Invited talk at the XXVIII International Symposium 
  on Physics in Collision, Perugia,  Italy, 25-28 June 2008.}}

\bigskip\bigskip


\begin{raggedright}  

{\it Vladimir Chekelian (Shekelyan)\index{Chekelian, V.}\\
Max-Planck-Institute f\"ur Physik \\
F\"ohringer Ring 6, 80805 Munich, Germany}
\bigskip\bigskip
\end{raggedright}

\begin{abstract}
Measurements of the longitudinal structure function $F_{L}$
at the $ep$ collider HERA are presented. They are 
derived from inclusive deep inelastic neutral current
$e^+p$ scattering cross section measurements 
based on data collected in 2007 with the H1 and ZEUS detectors
at a positron beam energy of 
$27.5$~GeV and proton beam energies of $920, 575$ and $460$~GeV. 
Employing the energy dependence of the cross sections,
$F_{L}(x,Q^{2})$ is measured in the range of negative four-momentum 
transfer squared $12 \leq Q^2 \leq 800~$GeV$^2$ and low Bjorken-$x$ 
$0.00028\leq x \leq 0.0353$. 
The measured longitudinal structure function 
is compared with higher order QCD predictions.
\end{abstract}

\section{Introduction}
In the past 15 years, the HERA experiments ~H1 and ZEUS
have extended the knowledge of the proton structure 
by two orders of magnitude towards high  
negative four-momentum transfer squared, $Q^2$,
and to small Bjorken-$x$. 
At the end of the HERA operation in 2007, 
dedicated $e^+p$ data were collected with
lower proton beam energy which allow to measure 
the longitudinal component of the proton structure. 
This measurement is directly sensitive to the
gluon contribution in the proton.
It is essential for completion of 
the deep inelastic scattering (DIS) program at HERA and 
for checks of the underlying perturbative Quantum Chromodynamics 
(QCD) framework used to determine parton distribution functions (PDFs).

The DIS neutral current (NC) $ep$ scattering cross section at low $Q^2$ 
can be written in reduced form as
\begin{equation}
 \sigma_r(x,Q^2,y) = \frac{d^2\sigma}{dxdQ^2} 
 \cdot \frac{Q^4 x}{2\pi \alpha^2 [1+(1-y)^2]}  
  =  F_2(x,Q^2) - \frac{y^2}{1+(1-y)^2} \cdot  F_L(x,Q^2)~.
       \label{sig}
  \end{equation}  
Here, $\alpha$ denotes the fine structure constant,
$x$ is the Bjorken scaling variable and $y$ is 
the inelasticity  of the scattering process
related to $Q^2$ and $x$ by $y=Q^2/sx$,
where $s$ is the centre-of-mass energy squared of the incoming 
electron and proton.

The cross section is determined by two independent structure functions,  
$F_2$ and $F_L$. They are related to
the $\gamma^* p$ interaction cross sections
of longitudinally and transversely polarised virtual photons,
$\sigma_L$ and $\sigma_T$, according to
$F_2 \propto (\sigma_L + \sigma_T)$ and $F_L \propto \sigma_L$,  
therefore  $0 \leq F_L \leq F_2$.
$F_2$ is the sum of the quark and anti-quark $x$ distributions
weighted by the electric charges of quarks squared and 
contains the dominant contribution to the cross section.
In the Quark Parton Model the value of 
the longitudinal structure function $F_L$ is zero, 
whereas in QCD it differs from zero
due to gluon and (anti)quarks emissions.
At low $x$ the gluon contribution to $F_L$ exceeds the quark contribution 
and $F_L$ is a direct measure of the gluon $x$ distribution.

The longitudinal structure function, or equivalently 
$R=\sigma_L/\sigma_T=F_L/(F_2-F_L)$, was measured
previously in fixed target experiments and found to be small  at 
large $x \geq 0.2$,  confirming 
the spin 1/2 nature of the constituent quarks in the proton.
From next-to-leading order (NLO) and NNLO\,\cite{nnlo} QCD analyses of the
inclusive DIS cross section data \cite{mrsw,cteq,alekhin},
and from experimental $F_L$ determinations by H1\,\cite{h1fl,h1pdf2k},
which used assumptions on the behaviour of $F_2$,
the longitudinal structure function $F_L$ at low $x$ is
expected to be significantly larger than zero.
A direct, free from theoretical assumptions, measurement of $F_L$ at HERA,
and its comparison with predictions derived from the
gluon distribution extracted from the $Q^2$ evolution of
$F_2(x,Q^2)$ thus represents a crucial test 
on the validity of the perturbative QCD framework at low $x$.

\section{Measurement Strategy}
The model independent measurement of $F_L$  
requires several sets of NC cross sections
at fixed $x$ and $Q^2$ but different $y$. This was 
achieved at HERA by variation of the proton beam energy.

The measurements of the NC cross sections by H1 and ZEUS
are performed using $e^+p$ data collected in 2007 
with a positron beam energy  $E_e=27.5$\,GeV and with
three proton beam energies: the nominal
energy $E_p=920$\,GeV, the smallest energy
of $460$\,GeV and an intermediate energy of $575$\,GeV.
The corresponding integrated luminosities are about
$46$\,pb$^{-1}$, $12$\,pb$^{-1}$ and $6$\,pb$^{-1}$.  

The sensitivity to $F_L$ is largest at high $y$ as 
its contribution to $\sigma_r$ is proportional to $y^2$.
The high reconstructed $y$ values correspond to
low values of the scattered positron energy, $E_e'$:
\begin{equation}
y = 1 - {E_e' \over E_e} {\sin}^2 (\theta_e /2)~, \; \; \; \;
{Q}^2 = {{E_e'}^2 {\sin} ^ 2 \theta_e \over 1-y}~,  \; \; \; \; 
x=Q^2/sy. \enspace
\end{equation}
The measurement in the high $y$ domain up to $y=0.90$ 
requires the measurement of the scattered positron down to 
$E_e' \approx 3$\,GeV.
Thus, one needs a reliable identification and reconstruction 
of events with a low scattered positron energy. 
Furthermore, small energy depositions caused by hadronic final state 
particles can also lead to fake positron signals.
The large size of this background, mostly 
due to the photoproduction process at $Q^2 \simeq 0$,  
makes the measurement at high $y$ especially challenging.

\subsection{H1 Analysis}
H1 performed two independent analyses, at medium\,\cite{h1flmed}
 and high\,\cite{h1flhigh} $Q^2$,
with the positrons scattered into the acceptance of 
the backward Spacal calorimeter (the polar angle 
range of the scattered positron $\theta_e \gtrsim 153^{\circ}$),
corresponding to $12 \leq Q^2\leq 90$\,GeV$^2$, and 
into the acceptance of the Liquid Argon calorimeter (LAr) 
($\theta_e \lesssim 153^{\circ}$), corresponding to
$35 \leq Q^2\leq 800$\,GeV$^2$.

The scattered positron is identified as a localised energy 
deposition (cluster) with energy $E_e'>3.4 (>3) $\,GeV in the Spacal (LAr).
The NC events are triggered 
on positron energy depositions in the Spacal or LAr calorimeters,
on hadronic final state energy depositions in the Spacal,
and using a new trigger hardware commissioned in 2006.
At small positron energies the Spacal trigger
is complemented by the central inner proportional chamber (CIP)
track trigger which reduces the trigger rate to an acceptable level.
The new trigger system includes the Jet Trigger, which
performs a real time clustering in the LAr, and the Fast Track Trigger
(FTT)\,\cite{FTT}, which utilizes on-line reconstructed tracks in the
central tracker (CT). 
The combined trigger efficiency reaches $97-98$\% at $E_e'=3$\,GeV and
$\approx 100$\% at $E_e'>6-7$\,GeV.

To ensure a good reconstruction of kinematical properties, 
the reconstructed event vertex 
is required to be within 35 cm around the nominal vertex position
along the beam axis. The primary vertex position
is measured using tracks reconstructed in the central tracker system.
The positron polar angle is determined by the positions of the
interaction vertex and the positron cluster in the calorimeter. 

The photoproduction background is reduced by demanding
a track from the primary interaction 
pointing to the positron cluster
with an extrapolated distance to the cluster below 6 (12) cm
in the medium (high) $Q^2$ analysis. 
In the medium $Q^2$ analysis the fake positron background
is reduced by the requirement of a small transverse size of the cluster
in the Spacal, $R_{log}$, which is estimated using a logarithmic energy 
weighted cluster radius, and by the requirement that the energy 
behind the cluster, measured in the hadronic part of the Spacal,
may not exceed 15\% of $E_e'$.
For $E_e'<6$\,GeV in the high $Q^2$ analysis the following additional 
requirements are applied: 
small transverse energy weighted radius of the cluster (Ecra $ < 4$ cm)
and matching between the energy of the cluster and
the track momentum ($0.7 < E_{e}'/P_{track} < 1.5$).

Further suppression of photoproduction
background is achieved by requiring longitudinal 
energy-momentum conservation $\Sigma_i(E_i-p_{z,i})>35$\,GeV,
where the sum runs over the energy and longitudinal momentum
component of all particles in the final state including the
scattered positron. 
For genuine, non-radiative NC events it is equal to 
2$E_e = 55$\,GeV.
This requirement also suppresses events with hard initial
state photon radiation. 
QED Compton events are excluded using a topological
cut against two back-to-back energy depositions in the 
calorimeters.

In addition, a method of statistical background subtraction 
is applied for the $E_p=460$ and $575$\,GeV data 
at high y ($0.38<y<0.90$ and $E_{e}'<18$\,GeV).
The method relies on the determination of
the electric charge of the positron
candidate from the curvature of the associated track.
Only candidates with right (positive) sign of electric charge
are accepted.
The photoproduction background events are about equally
shared between positive and negative charges.
Thus, by selecting the right charge the background is suppressed
by about a factor of two. 
The remaining background is corrected for
by statistical subtraction of background events with 
the wrong (negative) charge 
from the right sign event distributions.
This subtraction procedure requires a correction for a small 
but non-negligible charge asymmetry in the background events
due to enlarged energy depositions in the annihilation of anti-protons
in the calorimeters. 

The small photoproduction background for the $920$\,GeV data 
at $y < 0.5 (0.56) $ in the medium (high) $Q^2$ analysis 
is estimated and subtracted using a PHOJET (PYTHIA) simulation 
normalised to the photoproduction data tagged in the electron tagger
located downstream of the positron beam at 6 m.

\subsection{ZEUS Analysis}

ZEUS developed special triggers
to record events with low positron energy. The performance of these 
triggers was demonstrated in a ZEUS measurement of 
NC cross section at high y\,\cite{zeushighy}.

The scattered positron with energy $E_e'>6 $\,GeV is identified 
in the ZEUS rear calorimeter at radii above $\approx 28$ cm 
from the beam line.
The standard tracking in ZEUS is limited to 
the particle polar angles of $\theta \lesssim 154^{\circ}$.
To suppress large photoproduction background caused by neutral particles
(mostly photons from $\pi^{\circ}$ decays) 
it is crucial to extend this region. A new method is used to define a corridor
between the reconstructed vertex of the event and the position of
the positron candidate in the calorimeter. Counting hits in the
tracking detector within the corridor and comparing it to the
number of the traversed layers in the tracker 
allows to discriminate between charged 
and neutral particles up to $\theta \lesssim 168^{\circ}$.

The remaining background among positron candidates was estimated using 
the PYTHIA simulation of photoproduction events.
The MC sample was normalised using the rate of identified
photoproduction events with a fake positron candidate
in the main detector and a signal in the electron tagger
located downstream of the positron beam at 6 m.

To ensure a good reconstruction of kinematical properties
and to further suppress photoproduction background,  
the reconstructed event vertex 
is required to be within 30 cm around the nominal vertex position
and the measured value of $\Sigma_i(E_i-p_{z,i})$ 
to be between 42 and 65\,GeV.

\begin{figure}[htb]
\begin{center}
\epsfig{file=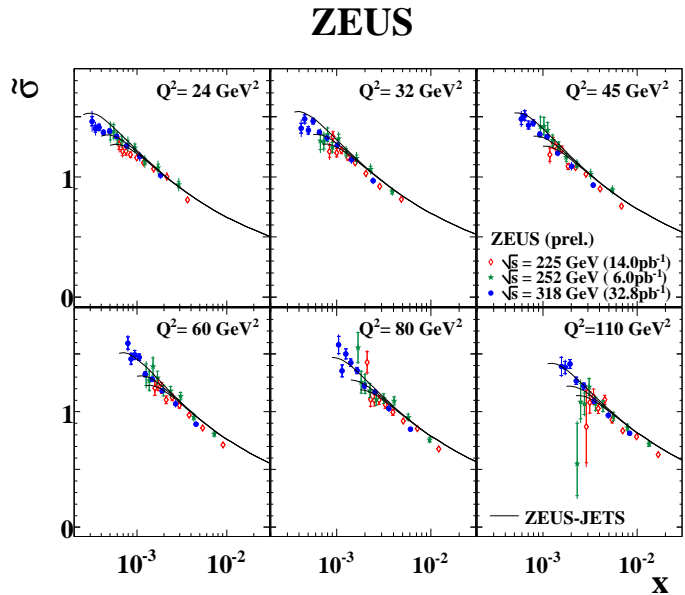,height=8.0cm}
\caption{
Reduced cross section measured by ZEUS at 
proton beam energies of 920, 575 and 460~GeV 
as a function of $x$ at fixed values of $Q^{2}$.
}
\label{fig1}
\end{center}
\end{figure}

\begin{figure}[htb]
\begin{center}
\epsfig{file=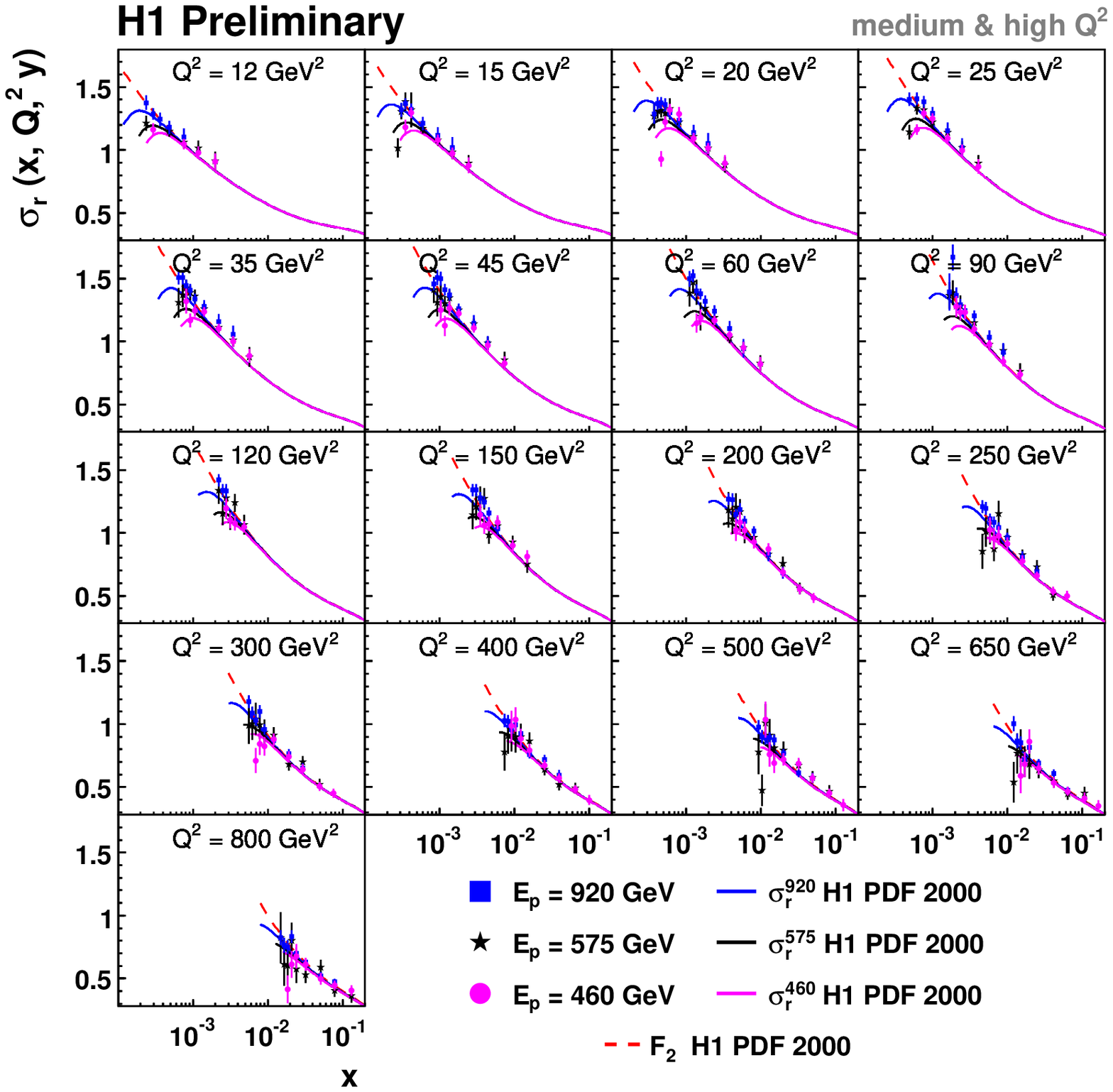,height=7.6cm}
\epsfig{file=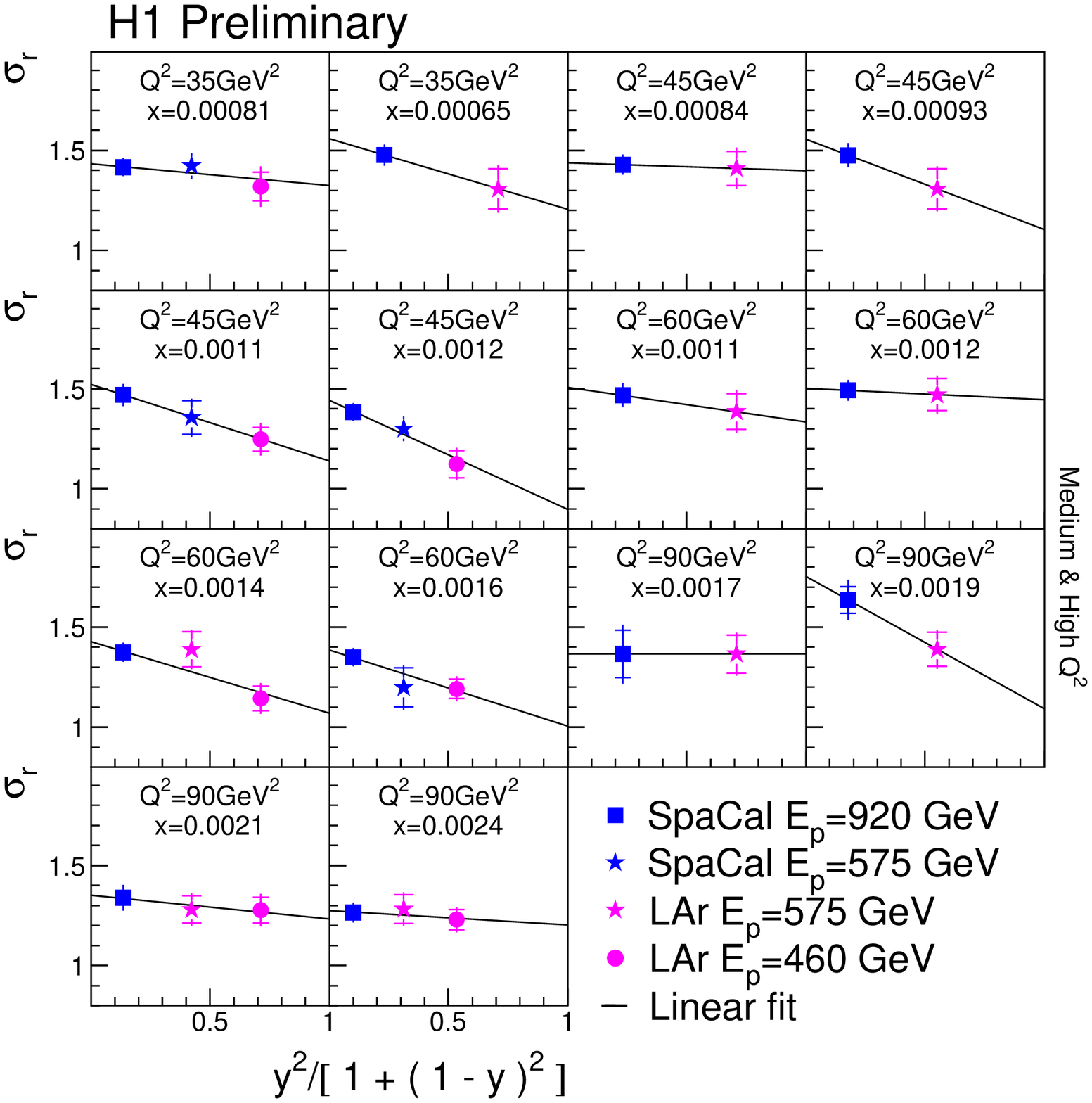,height=7.2cm}
\caption{
Reduced cross section measured by H1 at 
proton beam energies of 920, 575 and 460~GeV 
as a function of $x$ at fixed values of $Q^{2}$ (left)
and at fixed values of $x$ and $Q^{2}$ as a function of 
$y^2/[1+(1-y)^2]$
for measurements which include both the LAr and Spacal data (right).
The lines in the right figure show the linear fits used to determine 
$F_{L}(x,Q^2)$ for selected $Q^2$ and $x$.
}
\label{fig2}
\end{center}
\end{figure}

\section{HERA Results for {\boldmath $F_{L}(x,Q^2)$} }
The reduced NC cross sections measured by ZEUS\,\cite{zeusfl}
for three proton beam energies of $460$, $575$ and $920$\,GeV
are shown in figure\,\ref{fig1}
as a function of $x$ at fixed values of $Q^2$.
They cover the range $24 \leq Q^2\leq 110$\,GeV$^2$ 
and  $0.1\leq y \leq 0.8$ which
results in different coverage in $x$. 
The measurements are compared with 
the NLO QCD predictions based on the ZEUS-JETS QCD fit\,\cite{zeus-jet} 
taking into account the contribution from $F_L$
which causes a turn over of the expected cross section
at lowest $x$ values measured. 

The reduced NC cross sections measured by H1 at medium and high $Q^2$
($35 \leq Q^2\leq 800$\,GeV$^2$) in the range $0.1\leq y \leq 0.56$ 
for the $E_p=920$\,GeV data
and $0.1\leq y \leq 0.9$ for the $460$ and $575$\,GeV data
are shown in figure\,\ref{fig2} (left).
At $Q^2 \leq 25$~\,GeV$^2$ ($Q^2\geq 120$~\,GeV$^2$) the measurements 
are entirely from the  medium (high) $Q^2$ analysis.
In the intermediate $Q^2$ range $35 \leq Q^2\leq 90$\,GeV$^2$ 
the cross section is measured at $E_p=460 (920)$\,GeV in
the medium (high) $Q^2$ analysis
and for the $E_p=575$\,GeV data the cross section is obtained
either using the LAr or Spacal.
Small, 1-2\%, relative normalisation corrections 
to the measured cross sections 
at $E_p=460$, $575$ and $920$\,GeV, common for both analyses, 
are derived using measurements at low $y$ and
applied to the cross section points shown in the figure.
In this low $y$ region, the cross sections are determined by $F_2(x,Q^2)$
only, apart from a small correction for residual $F_{L}$ contribution.

\begin{figure}[htb]
\begin{center}
\epsfig{file=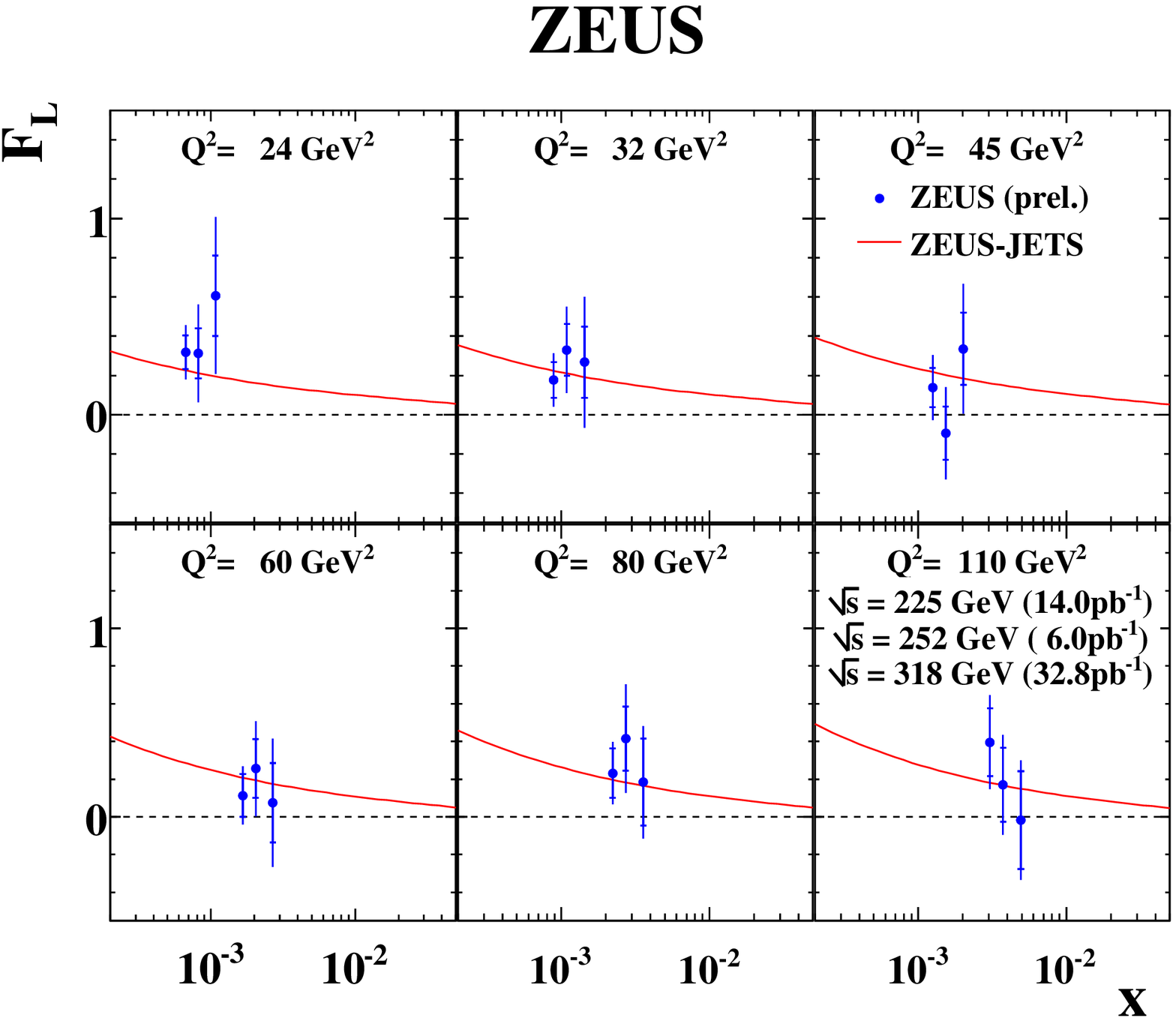,height=8.0cm}
\caption{
$F_{L}(x,Q^2)$ measured by ZEUS as 
a function of $x$ at fixed values of $Q^{2}$.
The inner and outer error bars are the statistical and total errors,
respectively.
The curve represents the NLO QCD prediction derived from
the ZEUS-JETS fit to previous ZEUS data.
}
\label{fig3}
\end{center}
\end{figure}

\begin{figure}[htb]
\begin{center}
\epsfig{file=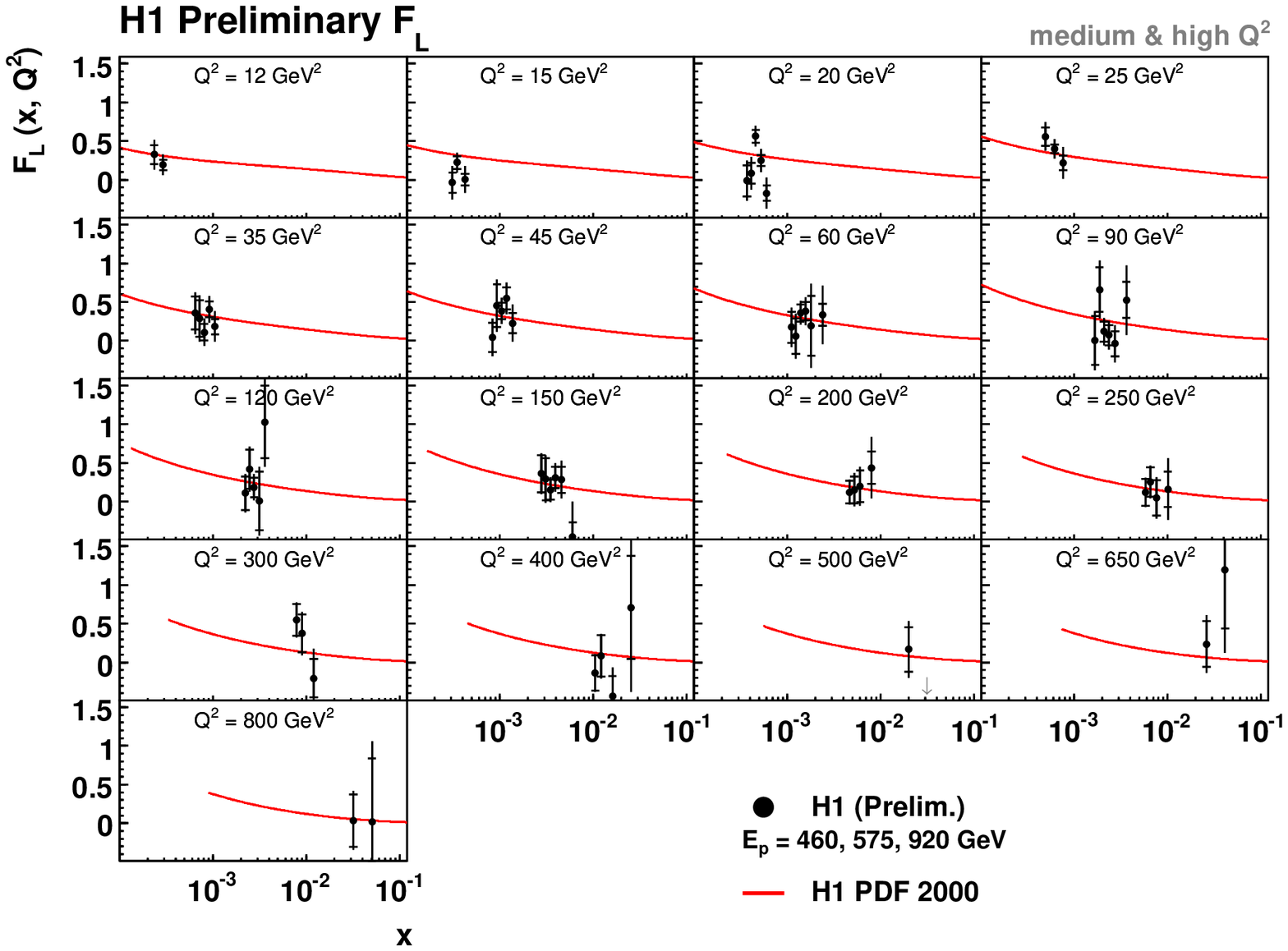,height=11.5cm}
\caption{
$F_{L}(x,Q^2)$ measured by H1 as 
a function of $x$ at fixed values of $Q^{2}$.
The inner and outer error bars are the statistical and total errors,
respectively.
The curve represents the NLO QCD prediction derived from
the H1 PDF 2000 fit to previous H1 data.
}
\label{fig4}
\end{center}
\end{figure}

The longitudinal structure function is extracted from 
the slope of the measured reduced cross section
versus $y^2/[1+(1-y)^2]$.
This procedure is illustrated for the H1 analysis
in figure\,\ref{fig2} (right)
for selected $Q^2$ and $x$ values, 
where both the LAr and Spacal measurements are available. 
The measurements are consistent with the expected linear dependence,
demonstrating consistency of the two independent
analyses, which utilize different detectors to measure 
the scattered positron.

The $F_L(x,Q^2)$ values are determined in straight-line fits 
to the $\sigma_r(x,Q^2,y)$.
The ZEUS result for $F_L(x,Q^2)$ is shown in figure\,\ref{fig3}.
The result is consistent with the expectation based 
on the ZEUS-JETS QCD fit\,\cite{zeus-jet}.

The H1 measurements of $F_L(x,Q^2)$ with statistical errors better 
than 10\% are shown in figure\,\ref{fig4}. The central $F_L(x,Q^2)$ 
values are determined in the fits using statistical and uncorrelated 
systematic errors added in quadrature, and statistical (total) $F_L$ 
errors - in the fits using statistical (total) errors.
The  uncertainty due to the relative normalisation of the cross sections 
is added in quadrature to the total $F_L(x,Q^2)$ error.
This uncertainty is estimated from the effect of a 1\% variation
of the normalisation 
of the $920$\,GeV cross section on the fit result.
The measurement of $F_L(x,Q^2)$ is limited to $Q^2$ and $x$ values 
where the total $F_L$ error is below $0.4$ ($1.1$) 
for $Q^2 \leq 35$ ($>35$)\,GeV$^2$.
The result is consistent with the NLO QCD prediction based on the
H1 PDF 2000 fit\,\cite{h1pdf2k} performed using 
previous H1 cross section data at nominal proton energy.

\begin{figure}[htb]
\begin{center}
\epsfig{file=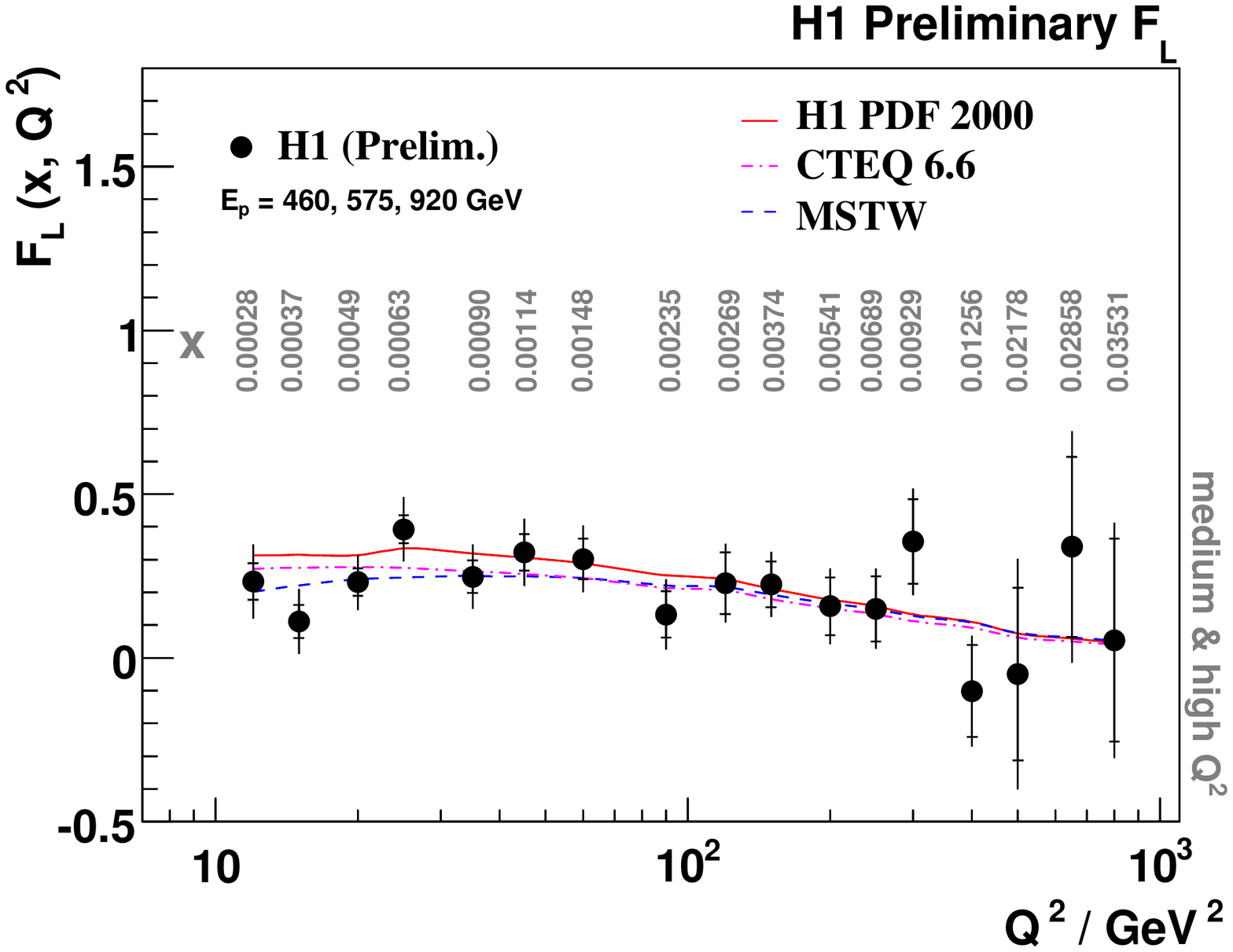,height=9.0cm}
\caption{
The H1 measurement of $F_{L}(Q^2)$ averaged over $x$ at fixed 
values of $Q^{2}$.
The resulting x values of the averaged $F_{L}$
are given in the figure for each point in $Q^{2}$.
The curves represent the QCD predictions.
}
\label{fig5}
\end{center}
\end{figure}
The H1 measurements of $F_L(Q^2)$ averaged over $x$ at fixed $Q^2$ 
are presented in  figure\,\ref{fig5}.
The average is performed using the total errors of individual measurements.
The overall correlated component for the averaged  $F_{L}$ 
is estimated to vary between 0.05 and 0.10.
The averaged $F_L$ is
compared with the H1 PDF 2000 fit\,\cite{h1pdf2k} and with
the expectations from global parton distribution
fit at NNLO (NLO) perturbation theory performed by the 
MSTW\,\cite{mrsw} (CTEQ \,\cite{cteq}) group
and from the NNLO QCD fit by Alekhin\,\cite{alekhin} 
(see also figures in \cite{h1flhigh}). 
Within the experimental
uncertainties the data are consistent with these predictions. 

\newpage
\section{Summary}

The H1 and ZEUS measurements of the longitudinal proton structure function 
in deep inelastic scattering at low $x$ are presented.   
The $F_L(x,Q^2)$ values are extracted from three sets of cross section
measurements at fixed $x$ and $Q^2$, but different
inelasticity $y$
obtained using three different proton beam energies at HERA. 
For the $Q^2$ range between $12$ and $800$\,GeV$^2$, 
the $F_L$ results are consistent at the current level of accuracy
with the DGLAP evolution framework of perturbative QCD at low $x$.



\end{document}